\documentclass{ws-procs9x6}

\begin{document}

\title{WIMPs search by exclusive measurements with thin multilayer 
NaI(Tl) scintillators (PICO-LON)}

\author{K. Fushimi$^*$, H.Kawasuso, K.Yasuda, Y.Kameda, 
N.Koori and S.Nakayama}

\address{Faculty of Integrated Arts and Sciences, The University of Tokushima\\
Tokushima city, Tokushima 770-8502, JAPAN\\
$^*$E-mail: kfushimi@ias.tokushima-u.ac.jp}

\author{K.Ichihara, M.Nomachi, S.Umehara}

\address{Department of Physics, Osaka University\\
Toyonaka city, Osaka 560-0047, JAPAN}
\author{R.Hazama}
\address{Graduate School and Faculty of Engineering, Hiroshima University\\
Higashi-Hiroshima city, Hiroshima 739-8527, JAPAN}
\author{S.Yoshida}
\address{Department of Physics, Tohoku University\\
Sendai city, Miyagi 980-8678, JAPAN}
\author{H.Ejiri}
\address{Department of Physics, International Christian University\\
Mitaka, Tokyo 181-8585, JAPAN}

\author{K.Imagawa and H.Ito}
\address{Horiba Co. Ltd.\\
Kyoto city, Kyoto 601-8510, JAPAN}
\begin{abstract}
The WIMPs search project PICO-LON has been started with multilayer 
thin NaI(Tl) crystals.
The thin (0.05cm) and wide area (5cm$\times$5cm) NaI(Tl) crystals was 
successfully developed.
The performances of thin NaI(Tl) scintillator was measured and 
they showed good energy resolution (20\% at 60keV) and 
good position resolution (20\% in 5cm$\times$5cm wider area).
\end{abstract}


\bodymatter

\section{Introduction}\label{aba:sec1}
Dark matter search are one of the most 
important subjects in nuclear- and particle-physics.
The particle candidates for cosmic dark matter is a key subject 
not only astrophysics but also particle physics since the particle 
candidates for dark matter is proposed by various models of 
the beyond standard model.

The components of the universe has been clearly understood by many cosmological
observations.\cite{WMAP,Lyalpha,2dF}.
Since, the most of the matter in the universe should be the cold dark matter, 
searching for WIMPs (Weakly Interacting Massive Particles) is 
quite important.
Recently, a direct empirical evidence for dark matter in a galaxy 
has been reported \cite{direct,direct2}.
The dark matter in the galaxy has become quite ensuring.
One of the most promising candidate for WIMPs is SUSY neutralino, which 
interacts with the matter via only weak interaction.

\begin{figure}[ht]
\epsfxsize=10cm
\epsfbox{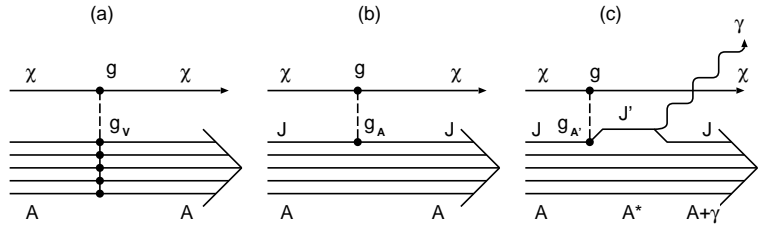}
\caption{
Feinman diagrams for WIMPs-nucleus interaction.
``A'' and ``$\chi$'' means the nucleus and the WIMPs.
(a) Spin independent elastic scattering (SI).
(b) Spin dependent elastic scattering (SD).
(c) Spin dependent inelastic scattering (EX).
}
\label{fg:feinman}
\end{figure}

The processes for WIMPs-nucleus interaction are schematically illustrated in 
Fig.\ref{fg:feinman}.
In Fig.\ref{fg:feinman}(a), the scattering amplitude is summed coherently 
over all the nucleon.
Thus the scattering cross section is coherently enhanced by the factor 
$A^{2}$,
where $A$ is the mass number of the target nucleus.
In Fig.\ref{fg:feinman}(b),  only one nucleon which carries the nuclear spin
contributes the cross section.
Thus the cross section is a few orders of magnitude smaller than that of 
the previous case. 
The cross section depends on the nuclear spin-matrix element 
$\lambda^{2}J(J+1)$.
and has a large ambiguity because 
this matrix element has a large model dependence for the heavy nuclei 
\cite{Ellis}.
In Fig.\ref{fg:feinman}(c), the target nucleus will be excited to the 
low lying excited state, which is followed by the gamma ray emission.
This process is arisen only if the WIMPs particle has enough kinetic energy.
In this case, the matrix element is measured precisely by the nuclear 
de-excitation\cite{Ellis,Inel1,Inel2}. 
Thus the model dependence of the cross section is much smaller than 
the cross section of the SD case.

The segmentation of the detector is shown to be the best way to 
enhance the position sensitivity\cite{FushimiJPSJ74}.
Coincidence measurements of nuclear recoils and $\gamma$ rays for the 
inelastic excitation of $^{127}$I enhanced the sensitivity using 
the highly segmented NaI(Tl) detector.
Recently, ionized atomic electrons and hard X rays following 
WIMPs-nuclear interactions have been shown to be useful for 
the exclusive measurement of
nuclear recoils from the elastic scatterings of WIMPs off nuclei
\cite{ejiri-X,ejiri-X2}.
On the other hand, the background events have their own characteristics of 
timing and spatial profiles.
Because the event rate due to the background is reduced by segmentation, 
a probability of the accidental coincidence of 
individual background events is vastly reduced.

It has been shown that piling up many thin scintillators enhances 
the sensitivity for WIMPs search\cite{FushimiJPSJ74}.
The highly sensitive detector system PICO-LON
(Planar Inorganic Crystals Observatory for LOw-background Neutr(al)ino)
has been developed.
It consists of many thin NaI(Tl) crystals whose thickness is 0.05cm.
Recently, PICO-LON-I which was made up three crystals of 
thin (0.05cm in thickness) and wide area (6.6cm$\times$6.6cm) 
NaI(Tl) crystals has been developed.
PICO-LON-II which was made up sixteen crystals of thin NaI(Tl) has been 
also developed.
In this report, we describe the excellent performance of a single 
plate of thin NaI(Tl) which is the foundation of PICO-LON system.

\section{The performance of thin NaI(Tl) scintillator}
The performance of the thin NaI(Tl) scintillator 
was measured by irradiating a thin NaI(Tl) crystal whose dimension was 
5cm$\times$5cm$\times$0.05cm
with low-energy $\gamma$ rays and X rays.
The scintillation photons were collected at the four edges of 
the NaI(Tl) crystal using four photomultiplier tubes (PMTs),
which were provided by Hamamatsu Photonics (R329-P).

The gamma rays and X rays were irradiated isotropically on the 
wider surface of the NaI(Tl) plate.
Each PMT output signal was individually input into four discriminators.
The threshold of the discriminators was set above 
that of the single-photoelectron signal;
the corresponding hardware energy threshold was 0.8keV.\@
The four PMT outputs were individually converted to digital data using 
a charge integrating analog-to-digital converter (RPC-022).
The total charge outputs of the PMTs were summed event-by-event 
using an off-line analyzer.

\begin{figure}[ht]
\begin{center}
\includegraphics[width=5.7cm]{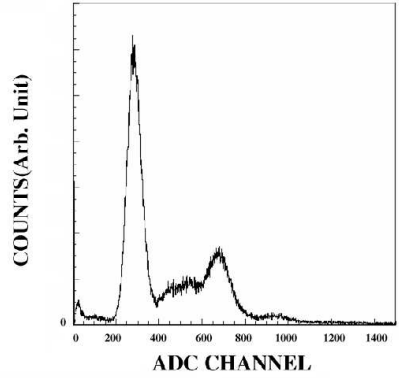}
\end{center}
\caption{
Pulse height spectrum of $^{133}$Ba.
}
\label{fg:spect}
\end{figure}

The resulting photon outputs are shown in Fig.\ref{fg:spect}.
In the pulse height spectrum of $^{133}$Ba, 
high-energy gamma rays of energies  
above 200keV were not clearly observed because the detector was too thin to 
absorb the gamma rays.
A photoelectric peak of 81keV and the corresponding X ray escape peak were 
observed at approximately 670ch and 500ch, respectively.
The prominent peak at approximately 300ch was due to the K-X rays of cesium.
Note that the small peak due to 
the low-energy L-X rays of cesium is observed approximately 30ch.
It is important in the search for WIMPs to be able to 
observe low energy, 
and the present results correspond to the energy threshold being 2keV.\@
The results showed that the thin NaI(Tl) scintillator displays great 
promise in the search for WIMPs.

The energy resolutions at FWHM (Full Width at Half Maximum) were calculated 
from the peaks and are shown in Table \ref{tb:res}.
\begin{table}[ht]
\tbl{
Energy resolutions for low-energy photons at FWHM.\@
The calculated number of photoelectrons (P.E.) is listed in the fourth column.
\cite{FushimiJPSJ75}}
{\begin{tabular}{lrrr} \hline
Source   & Energy (keV) & $\Delta E/E$ (FWHM) & \# of P.E.\\ \hline
$^{133}$Ba  & 81  & 0.17 & 197\\
$^{241}$Am  & 60  & 0.20 & 143\\
$^{133}$Ba  & 31  & 0.28 & 71\\ 
$^{57}$Co   & 14.4 & 0.40 & 35\\ \hline
\end{tabular}}
\label{tb:res}
\end{table}

The scintillation output has good linearity up to 120keV.\@
From the pulse height spectra, the low energy threshold was found to be
2$\sim$3keV.\@
The energy equivalent to a single photoelectron was also 
calculated using the photoelectron number $N$.
The energy threshold of approximately 2$\sim3$keV corresponds to 4$\sim$5 
photoelectrons.
The results showed an excellent performance that is in accordance with 
the required performance for the advanced stage of the experiment.

The position resolution for the thinner directions is as good as 0.05cm 
because of the segmentation of the detector.
Moreover, the position resolution for other directions was tested.
Because the largest area has dimensions of 5cm$\times$5cm,
good position information in the wider area enhances detector sensitivity.
Position information was obtained by analyzing the ratio of 
the number of photons collected on the opposite sides of the detector.
Precise position information on the largest area of the thin NaI(Tl) 
scintillator is important to ascertain the property of the events.
By piling up the thin NaI(Tl) scintillator, the tracking of radiation 
such as cosmic rays and the multiple Compton scattering of photons 
is reconstructed precisely.

A collimator for low-energy $\gamma$ rays was made of 1cm-thick lead brick
with nine holes with a diameter of 2mm was used for position measurement.
An $^{241}$Am source was placed at the top of 
each hole.
Position determination analysis was performed using 60keV gamma 
rays from $^{241}$Am.
The position resolution was calculated to be approximately 1cm in FWHM
\cite{FushimiJPSJ75}.

\section{Estimated sensitivity for WIMPs}
\begin{figure}[ht]
\begin{center}
\includegraphics[width=5.7cm]{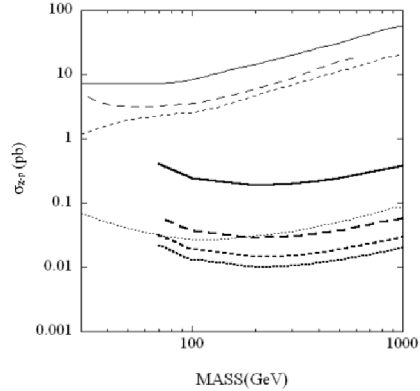}
\end{center}
\caption{
Thick lines are the expected sensitivity for SD type WIMPs.
The solid, long-dashed, sort-dashed and dotted lines are the 
expected sensitivity by 16, 256, 1024 and 2176 modules of NaI(Tl) array.
Thin lines are results of CRESST (solid line), DAMA LXe (long-dashed line),
ELEGANT V at OTO (short-dashed line) and the expected sensitivity by 
NAIAD (dotted line).\cite{FushimiJPSJ74}
}
\label{fg:sens}
\end{figure}

The case of inelastic scattering the nuclear spin-matrix element
is experimentally deduced from the nuclear transition probability,
consequently, the precise exclusion plot with small model dependence 
is obtained.
The estimated sensitivity for SD type WIMPs is shown in Fig.\ref{fg:sens}.
It is shown that the high sensitivity is expected by the small amount of 
NaI(Tl) crystal.

\section{Future prospects}
Test experiment with the three-layer NaI(Tl) detector (PICO-LON-I) has 
been performed at surface laboratory at Tokushima.
The PICO-LON detector will be installed into Oto Cosmo Observatory in 
the south of Nara prefecture where is 150km east from Tokushima and 
100km south from Osaka.

Oto Cosmo Observatory is covered with thick rock whose thickness is 
about 1200m.w.e.\cite{yoshida}.
Thus the flux of cosmic ray is reduced by a factor of $10^{-5}$\cite{OTO}.
The most serious origin of low energy background is $^{222}$Rn in the air.
Since the tunnel is opened at both ends and strong wind is running always, 
the concentration of Rn in the tunnel is the same as the one out of the tunnel.
The measured Rn concentration in the air in the laboratory was 
measured by highly sensitive Rn monitor which can measure 
above 5mBq/m$^{3}$.

PICO-LON-I and PICO-LON-II will be installed into the shield with the 
10cm thick OFHC(Oxygen Free High Conductive Copper) and 15cm thick old lead, 
which was used for the shield of ELEGANT V.
\bibliographystyle{ws-procs9x6}
\bibliography{fushimi-idm}
\end{document}